\documentstyle[prb,aps,eqsecnum,epsf]{revtex}
\newcommand{\be}{\begin{equation}}
\newcommand{\ee}{\end{equation}}
\newcommand{\bea}{\begin{eqnarray}}
\newcommand{\eea}{\end{eqnarray}}
\newcommand{\pas}{/ \kern-0.55em\partial}
\newcommand{\As}{/ \kern-0.55em A}
\newcommand{\Bs}{/ \kern-0.69em B}
\newcommand{\Cs}{/ \kern-0.55em C}

\begin{document}
\title{Unity of Fundamental Interactions}
\author{Ramchander R. Sastry}
\address{Center for Particle Physics,\\
 University of Texas at Austin, \\
  Austin, Texas 78712-1081.}
\date{\today}
\maketitle

\large %
\baselineskip=23.5pt plus .5pt minus .2pt
\begin{abstract}
The vector representation of the linearized gravitational
field (the graviton field) or the so called quantum
gravitodynamics which describes the motion of masses in a weak
gravitational field is employed to understand the unity of the
four known interactions.  We propose a gauge group
$SU(3)\times SU(2)\times U(1) \times U(1)$
for such a unified field theory.  In this paper we study the
$SU(2)\times U(1) \times U(1)$ sector of the theory and in analogy to
the electroweak mixing angle we define a gravitoweak mixing angle.
The unified gauge field theory predicts the existence of three massive
vector bosons, the $Y^{\pm}$ and the $X^0$, and two massless vector
bosons, the photon and the graviton (in its vector representation).
We determine the mass spectrum of the $Y^{\pm}$ and the $X^0$ and predict
a modification to the fine structure constant under unified field conditions.
Furthermore, we briefly discuss the implications of the extended object formulation for the gauge hierarchy problem.  

\end{abstract}

\normalsize
\baselineskip=21.6pt plus .5pt minus .2pt
\section{Introduction}
The quantum mechanics of extended objects \cite{sastry1} and its
infinite dimensional generalization, namely, the quantum field
theory of extended objects, in particular $\phi^6$ scalar field
theory \cite{sastry2}, quantum electrodynamics with the Pauli term, 
\cite{sastry3} and quantum gravitodynamics \cite{sastry4} have been
presented by the author.  In quantum gravitodynamics, the author
develops an approach to understanding the response of of a lepton to a 
weak (linearized) gravitational field by making use of the vector
representation of the linearized gravitational field \cite{wald}.
In the covariant perturbation theory approach, the quantum theory of
gravity is rendered finite by making use of a Euclidean, retarded,
graviton propagator given by:
\be
\label{prop}
\left(\delta{\mu\rho}\delta{\nu\sigma} + \delta{\mu\sigma}\delta{\nu\rho} 
- \delta{\mu\nu}\delta{\rho\sigma}\right)\frac{e^{-k^2/m^2}}{k^2}
\ee
where $\frac{1}{m}$ is the graviton Compton wavelength given by
 $6.7 \times 10^{-4}R$ where $R = c/H$ is the ``Hubble radius'' of the
 universe and $H$ is the Hubble constant.  The graviton propagator is defined
in the linear approximation since the notion of mass and spin of a field
requires the presence of a flat background metric $\eta_{\mu\nu}$ which one
does not have in the full theory.  The full theory of general relativity
may then be viewed as that of a graviton field which undergoes a nonlinear
self-interaction.  The propagator in Eq.~(\ref{prop}) will render such
a full theory finite to all orders.  It is the discovery of this propagator which motivates us to study the possibility of unifying the graviton field with
 the existing electroweak theory.  It is known that
 linearized gravity predicts that the motion of masses produces magnetic
 gravitational effects very similar to electromagnetism \cite{wald}.
 The effective interaction between the
electron and the graviton field can be understood in the vector representation
where we make use of a propagator with the functional dependence given in
Eq.~(\ref{prop}) but with suitable vector indices.  The author has calculated
the order $\alpha$ correction to the magnetic gravitational moment by using
such a propagator \cite{sastry4}.  Therefore, we are motivated to propose
a gauge group $SU(3)\times SU(2)\times U(1) \times U(1)$ for the unified field
theory which incorporates the strong force, the weak and electromagnetic
interaction, and the graviton field.  In this paper, we focus on the 
$SU(2)\times U(1) \times U(1)$ sector of the gauge theory.  The feasibility
of such a gauge structure and its implications for the existence of massive
vector bosons, the $Y^{\pm}$ and the $X^0$, and the determination of their mass
spectrum are studied in this paper.  We also predict a modification to the
fine structure constant under unified field conditions.  Furthermore, the consequences of the extended object formulation for the gauge hierarchy problem are examined.
 
\section{$SU(2)\times U(1)\times U(1)$}

Let us consider the electronic-type lepton fields which consist of only
the left- and right-handed parts of the electron field e:
\be
e_L = \frac{1}{2}(1 + \gamma_5)e,\,\,\,\,\, e_R = \frac{1}{2}(1 - \gamma_5)e
\ee
and a purely left-handed electron-neutrino field $\nu_{eL}$:
\be
\gamma_5\nu_{eL} = \nu_{eL}.
\ee
In any representation of the gauge group, the fields must all have the same 
Lorentz transformation properties, so the representations of the gauge group
must divide into a left-handed doublet $(\nu_{eL},e_L)$ and a right handed
singlet $e_R$.  Thus, the largest possible gauge group is then
\be
SU(2)\times U(1)\times U(1)
\ee
under which the fields transform as
\be
\delta\left(\matrix{\nu_e\cr e}\right) = i\left[\vec{\epsilon}\cdot \vec{t
} + \epsilon_L t_L + \epsilon_R t_R\right]\left(\matrix{\nu_e\cr e}\right)
\ee
where the generators are
\bea
\label{t}
\vec{t} &=& \frac{g}{4}(1 + \gamma_5)
\left\{\left(\matrix{0&1\cr 1&0\cr}\right),
\left(\matrix{0&-i\cr i&0\cr}\right),
 \left(\matrix{1&0\cr 0&-1\cr}\right)\right\},\\
t_L &\propto& (1 + \gamma_5)\left(\matrix{1&0\cr 0&1\cr}\right),\\
t_R &\propto& (1 - \gamma_5)\\
\eea
with $g$ an unspecified constant.  It will be convenient instead of $t_L$
and $t_R$ to consider the generators
\be
\label{y}
y = g'\left[\frac{(1 + \gamma_5)}{4}\left(\matrix{1&0\cr 0&1\cr}\right) 
+ \frac{(1 - \gamma_5)}{2}\right]
\ee
and
\be
\label{ne}
n_e = g''\left[\frac{(1 + \gamma_5)}{2}\left(\matrix{1&0\cr 0&1\cr}\right)
      +\frac{(1 - \gamma_5)}{2}\right],
\ee
where $g'$ and $g''$ are unspecified constants like $g$.  The generator $y$
(the hypercharge) appears along with $t_3$ (the isospin operator) in a linear
combination to define the charge $q$ of the pair $(\nu_{eL},e_L)$:
\be
\label{q-eqn}
q = e\left(\matrix{0&0\cr 0&-1\cr}\right) = \frac{e}{g}t_3 - \frac{e}{g'}y.
\ee
Also, $n_e$ is the electron-type lepton number and it defines the mass of 
the left-handed pair $(\nu_{eL},e_L)$ and the right-handed singlet $e_R$:
\be
\label{m-eqn}
m_{ab} = m\left(\matrix{1&0\cr 0&1\cr}\right) = \frac{m}{g''}n_e
\ee
where $m$ is the electron mass.  Thus, the charge couples to the electromagnetic
 field and the mass (in geometrized units) couples to the weak gravitational field.
 We want to include charge changing
weak interactions (like beta decay), electromagnetism, and the graviton field in 
our theory, so we will assume there are gauge fields $\vec{A}^{\mu}$, $B^{\mu}$,
and $C^{\mu}$ coupled to $\vec{t}$, $y$, and $n_e$ respectively.  Before we include
the graviton field in our theory we must ensure that it satisfies the stringent
limits on long range forces that would be produced by a massless gauge field
coupled to $n_e$ \cite{lee}.  Since the gravitational interaction is much weaker than the 
weak or electromagnetic interactions we are free to include a gauge field $C^{\mu}$
with strength $g''$ coupled to $n_e$.  The gauge group is then
\be
G = SU(2)_L\times U(1)\times U(1)
\ee
where the generators $\vec{t}$, $y$, and $n_e$ are given by Eq.~(\ref{t}),
Eq.~(\ref{y}), and Eq.~(\ref{ne}) respectively.  The most general gauge-
invariant and renormalizable Lagrangian that involves gauge-fields and
electronic leptons is
\bea
{\cal L}_{YM} + {\cal L}_{LG} + {\cal L}_e =  
-\frac{1}{4}\left(\partial_{\mu}\vec{A_{\nu}}
 -  \partial_{\nu}\vec{A_{\mu}}
 + g\vec{A_{\mu}}\times \vec{A_{\nu}}\right)^2\\ \nonumber
 -\frac{1}{4}\left(\partial_{\mu}\vec{B_{\nu}}
 - \partial_{\nu}\vec{B_{\mu}}\right)^2
-\frac{1}{4}\left(\partial_{\mu}\vec{C_{\nu}}
 - \partial_{\nu}\vec{C_{\mu}}\right)^2\\ \nonumber
 - \overline l\left(\pas - i\vec{\As}\cdot \vec{t} - i\Bs y - i\Cs n_e\right)l.
\eea
The coupling constants $g$ and $g'$ are to be adjusted so that the gauge fields
$\vec{A}^{\mu}$, $B^{\mu}$, and $C^{\mu}$ coupled to these generators are canonically
normalized.  Now, of these five gauge fields coupled to $\vec{t}$, $y$, and $n_e$,
 only two linear combinations, the electromagnetic field $A_{\mu}$, and
the graviton field (vector representation) $A_{\mu}^G$ are actually massless.
  We therefore must assume that $SU(2)_L\times U(1)\times U(1)$ is 
 spontaneously broken into $U(1)_{em}\times U(1)_{gravity}$ with generators
given by the hypercharge $y$ and the electron-type lepton number $n_e$.
  The details of the symmetry-breaking mechanism will be considered a 
 little later.  However, whatever this mechanism may be, we know that
the canonically normalized vector fields corresponding to particles of 
spin one and definite mass consist of one field of charge $+e$ with 
mass $m_Y$
\be
Y^{\mu} = \frac{1}{\sqrt{2}}\left(A_1^{\mu} + iA_2^{\mu}\right)
\ee
and another of charge $-e$ and the same mass
\be
Y^{\mu^*} = \frac{1}{\sqrt{2}}\left(A_1^{\mu} - iA_2^{\mu}\right)
\ee
and three electrically neutral fields of mass $m_X$, zero, and 
zero respectively given by orthonormal linear combinations of 
$A_3^{\mu}$, $B^{\mu}$, and $C^{\mu}$:
\bea
X^{\mu} &=& \cos\phi A_3^{\mu} + \sin\phi B^{\mu}\\
A^{\mu} &=& -\cos\theta\sin\phi A_3^{\mu} + \cos\theta\cos\phi B^{\mu}
 + \sin\theta C^{\mu}\\
A_G^{\mu} &=& \sin\theta\sin\phi A_3^{\mu} - \sin\theta\cos\phi B^{\mu}
 + \cos\theta C^{\mu}
\eea
 where $\phi$ is the electroweak mixing angle (the Weinberg angle) and $\theta$ is the 
gravitoweak mixing angle.  These linear combinations employ the Euler
 angles for a transformation from space axes to body coordinates with
 the third rotation set to zero.  In this theory, the third rotation
 is set to zero because both the electromagnetic and graviton
 fields are massless $U(1)$ gauge fields and two $U(1)$'s are 
 independent of each other.  Hence, the mixing angle between the
 electromagnetic and graviton fields (the electrogravity angle)
 is zero.  The electromagnetic field mixes with the weak interaction
via the Weinberg angle and the weak interaction in turn mixes with
the graviton field via the gravitoweak mixing angle.
By making use of the inverse transformation back to space axes
we have:
\bea
\label{comb}
A_3^{\mu} &=& \cos\phi X^{\mu} - \cos\theta\sin\phi A^{\mu}
 + \sin\theta\sin\phi A_G^{\mu}\\
B^{\mu} &=& \sin\phi X^{\mu} + \cos\theta\cos\phi A^{\mu}
 - \sin\theta\cos\phi A_G^{\mu}\\
C^{\mu} &=& \sin\theta A^{\mu}  + \cos\theta A_G^{\mu}
\eea
In the limit as the gravitoweak angle $\theta$ goes to zero,
 we recover the linear combinations necessary to generate the
 electroweak mass spectrum \cite{weinberg}.  
In the above linear combinations we observe that the massive
fields $Y_{\pm}^{\mu}$
 and $X^{\mu}$ are specified entirely in terms of the gauge fields
 $\vec{A}^{\mu}$ and $B^{\mu}$.  Since the electrogravity mixing angle is zero,
 spontaneous symmetry breaking, which generates the vector meson term, occurs only
 in the electroweak sector of the theory.  However, the coupling constants $g$
 and $g'$ of the electroweak sector are specified in terms of the coupling
 constant $g''$ of the gravity sector as shown below.  Thus, the spontaneous
 symmetry breaking of $SU(2)_L\times U(1)\times U(1)$ into  $U(1)\times U(1)$
 will generate two massless particles, namely, the photon and the graviton.
Now, the generators of the unbroken symmetries, which are here electromagnetic
and gravitodynamic gauge invariance are given by a linear combination of
generators in which the coefficients are the same as the coefficients of 
the canonically normalized gauge fields coupled to these generators \cite{weinberg}.
Inspecting Eqs.(\ref{comb}) shows that
\bea
q &=& -\cos\theta\sin\phi\,t_3 + \cos\theta\cos\phi\,y,\\
m_{ab} &=& \cos\theta\,n_e.
\eea
Comparing this with Eq.(\ref{q-eqn}) and Eq.(\ref{m-eqn}) gives then
\be
g = \frac{-e}{\cos\theta\sin\phi}, \,\,\,\, g' = \frac{-e}{\cos\theta\cos\phi},\,\, 
\,\,g'' = \frac{m}{\cos\theta}.
\ee
To complete the theory, we must now make some assumption about the mechanism
of symmetry breaking.  This mechanism must give masses not only to the $Y^{\pm}$
and $X^0$, but to the electron as well.  Thus, we assume a  `Yukawa' coupling
\be
{\cal L}_{\phi} = -G_e\overline{\left(\matrix{\nu_e\cr e\cr}\right)} = 
\left(\matrix{\phi^+\cr \phi^0\cr}\right)e_R + H.c.,
\ee
where $(\phi^+,\phi^0)$ is a doublet on which the $SU(2)_L\times U(1)$
generators are represented by the matrices:
\bea
\vec{t}^{(\phi)} &=& \frac{g}{2}\left\{\left(\matrix{0&1\cr 1&0\cr}\right),
\left(\matrix{0&-i\cr i&0\cr}\right), \left(\matrix{1&0\cr 0&-1\cr}\right)\right\},\\
y^{(\phi)} &=& \frac{-g'}{2}\left(\matrix{1&0\cr 0&1\cr}\right)
\eea
so that the charge matrix is
\be
q = e\left(\matrix{1&0\cr 0&0\cr}\right) = \frac{e}{g}t_3^{(\phi)} -
 \frac{e}{g'}y^{(\phi)}.
\ee
The most general form of the gauge-invariant term involving scalar and gauge
fields consistent with the $SU(2)_L\times U(1)$ sector of the theory is:
\be
\label{l-eqn}
{\cal L}_{\phi}
 = \frac{-1}{2}\Bigl|\left(\partial_{\mu} -
 i\vec{A}\cdot\vec{t}^{(\phi)} -
 iB_{\mu}y^{(\phi)}\right)\phi\Bigr|^2
- \frac{\mu^2}{2}\phi^{\dagger}\phi -
 \frac{\lambda}{4}(\phi^{\dagger}\phi)^2                                    
\ee
where $\lambda > 0$ and 
\be
\phi =\left(\matrix{\phi^+\cr \phi^0}\right).
\ee
 For $\mu^2 < 0$, there is a tree-approximation vacuum expectation value at 
the stationary point of the Lagrangian
\be
\langle\phi\rangle \langle\phi^{\dagger}\rangle = v^2 = |\mu^2|/\lambda
\ee
In unitarity gauge the vacuum expectation values of the components of $\phi$
are
\be
\langle\phi^+\rangle = 0, \,\,\,  \langle\phi^0\rangle = v > 0.
\ee
The scalar Lagrangian Eq.~(\ref{l-eqn}) then yields a vector meson mass
term of the form
\be
\frac{-v^2g^2}{4}Y_{\mu}^{\dagger}Y_{\mu} 
- \frac{v^2}{8}\left(g^2 + g'^2\right)X_{\mu}X^{\mu},
\ee
where
\bea
\frac{g}{g''}&=& \frac{-e/m}{\sin\phi},\\
\frac{g'}{g''} &=& \frac{-e/m}{\cos\phi},\\
g'' &=& \frac{m}{\cos\theta}.
\eea
Here, $\phi$ is the Weinberg angle and $\theta$ is the gravitoweak mixing angle.
We see that the photon mass is zero corresponding to an unbroken gauge symmetry
$U(1)_{em}$ and the graviton mass is also zero corresponding to another unbroken
gauge symmetry $U(1)_{gravity}$ while the $Y^{\pm}$ and $X^0$ have the masses
\be
m_Y = \frac{v|g|}{2},\,\,\,\,\,\,\,\,\,   m_X = \frac{\sqrt{g^2 + g'^2}}{2}.
\ee
Now, consider the relation \cite{weinberg}
\be
g^2/m_Y^2 = 4\sqrt{2}G_F
\ee
where $G_F = 1.16639(2)\times 10^{-5}\mbox{GeV}^{-2}$ is the Fermi constant.  This
relation is obtained by comparing the effective interaction between low
energy, $e$-type and $\mu$-type leptons with the effective 'V-A' theory which
is known to give a good description of muon decay.  This allows an immediate
determination of the vacuum expectation value as
\be
v = \frac{2m_Y}{g} = 247 \mbox{GeV}.
\ee
By making use of the known value of the electroweak mixing angle given by
$\sin^2\phi = 0.23$, we can determine the masses of $Y^{\pm}$ and $X^0$
in terms of the gravitoweak mixing angle as:
\bea
\label{mass}
m_Y &=& \frac{e_{\mu}v}{2|\cos\theta||\sin\phi|}
 = \frac{80.2 \mbox{GeV}}{|\cos\theta|}\\
\label{mass2}
m_X &=& \frac{e_{\mu}v}{2|\cos\theta||\sin2\phi|}
 = \frac{91.3 \mbox{GeV}}{|\cos\theta|}
\eea
where $e_{\mu}$ is the electric charge defined at a sliding scale $\mu$ 
comparable to the energies of interest.  We observe that as $\theta\rightarrow 0$
we regain the $W$ and $Z$ boson masses which is a result we expect.  Thus, a 
unified field theory predicts the existence of massive vector bosons $Y^{\pm}$
and $X^0$ with the mass spectrum given in Eqs.(\ref{mass})-(\ref{mass2}). 
If we express the
covariant derivative in Eq.~(\ref{l-eqn}) in terms of the mass eigenstate fields
$Y_{\mu}^{\pm}$, $X^{\mu}$, and $A_{\mu}$ we find that the coefficient of the 
electromagnetic interaction is not the electron charge $e$, but rather the effective
electron charge $e'$
\be
e' = \frac{e}{|\cos\theta|} = \frac{gg'}{\sqrt{g^2 + g'^2}}.
\ee
We observe that $e' \geq e$ with equality being achieved when the gravitoweak
mixing angle $\theta$ is zero.  The mixing between the weak interaction and the graviton
field causes an increase in the electromagnetic coupling strength.  This is because
the electromagnetic coupling is a function of the gauge couplings $g$ and $g'$
which have a dependence on $\theta$.  If $\alpha_G$ is the fine structure constant of an electron in the unified field, then we have:
\be
\frac{\alpha_G}{\alpha}  = \frac{1}{\cos^2\theta}
\ee 
implying that the fine structure constant suffers a modification.  This would mean
that if we were to measure the Lamb shift under unification conditions,
the correction to the $g$-factor of the electron would be
\be
a_e = \frac{\alpha_G}{2\pi} = \frac{0.0011597}{\cos^2\theta}.
\ee

\section{The Gauge Hierarchy Problem}
We begin with the reasonable observation that if  $SU(2)\times U(1)$ is broken by the 
vacuum expectation value of an elementary scalar field, then that scalar field should be part of the grand unification.  In order to produce a vacuum expectation value of the right size to give the observed $W$ and $Z$ boson masses, the Higgs scalar field must obtain a negative mass term of the size \cite{peskin}
\be
-\mu^2 \sim -(100 GeV)^2.
\ee
Now, the mass term can be expressed in terms of the vacuum expectation value $v$ as
\be
|\mu^2| = \lambda v^2
\ee
where $\lambda$ is the renormalizable coupling in $(\phi^{\dagger}\phi)^2$ charged scalar field theory.  Therefore, the $(mass)^2$ receives additive renormalizations.  In a theory with a
cutoff scale $\Lambda$, $\mu^2$ can be much smaller than $\Lambda^2$
 only if the bare mass of the scalar field is of the order $-\Lambda^2$
 and this value is canceled  down to $-\mu^2$ by radiative corrections. 
If our theory of nature contains very large scales of grand unification,
 then the appropriate value for $\Lambda$ is $10^{16}$ GeV or larger and
 it would require bizarre cancellations in the renormalized value of $\mu^2$.  Thus, the Higgs boson mass is very small compared to the grand unification
 scale.  It is a mystery as to why the $(mass)^2$ of the Higgs boson has
 a value $28$ orders of magnitude or more below its natural value and this
 question is referred to as the {\it gauge hierarchy problem}.  However,
 at grand unification energy scales the contributions of hitherto
 nonrenormalizable terms such as the Pauli term become significant
 \cite{sastry3}.  The description of quantum electrodynamics with the
 Pauli term necessitates the introduction of the quantum field theory
 of extended objects in which the finite extent of a particle defined
 via its Compton wavelength is incorporated into the field structure
 and leads to a finite interaction.  Since hitherto nonrenormalizable
 terms become important at grand unification scales, it would be more
 correct if we consider $SU(2)\times U(1)$ to be broken by the vacuum
 expectation value of a hitherto nonrenormalizable
 $(\phi^{\dagger}\phi)^3$ scalar field which can be rendered finite
 in the extended object formulation \cite{sastry2}.  The coupling
 $\lambda$ now becomes a finite coupling and the $(mass)^2$ does
 not receive additive renormalizations.  Consider the potential
\be
V(\phi) = -\mu^2(\phi^{\dagger}\phi) + \lambda(\phi^{\dagger}\phi)^3
\ee
which has a tree-approximation vacuum expectation value at
\be
\langle\phi\rangle \langle\phi^{\dagger}\rangle =\left(|\mu^2|/\lambda\right)^{\frac{1}{2}}
\ee
implying that
\be
|\mu^2| = \lambda v^4
\ee
where $\lambda$ is now a finite coupling.  Therefore, we can now expect the Higgs boson mass to be of the order of $100$ GeV without any conceptual difficulty.

\section{Conclusion}
The general gauge group $SU(3)\times SU(2)\times U(1) \times U(1)$ appears to describe the four known interactions in a consistent fashion.  We are able to predict the existence of gauge bosons $Y^{\pm}$ and $X^{0}$ for the $SU(2)\times U(1)\times U(1)$ sector of this unified theory and determine mass spectrum of the gauge bosons.  We have also shown
that the fine structure constant is modified under unified field conditions.
In addition, a possible resolution of the gauge hierarchy problem has been discussed.
The results of this paper need to be subjected to experimental tests.



\end{document}